\documentclass[prd,preprintnumbers,
twocolumn,
eqsecnum,floatfix,letterpaper,superscriptaddress,nofootinbib]{revtex4}
\usepackage[margin=1.0in]{geometry}
\usepackage{color}
\usepackage{amsmath,amssymb,graphicx}
\usepackage{bm}
\usepackage{times}
\usepackage{microtype}
\usepackage{booktabs}
\usepackage{subfigure}
\usepackage[normalem]{ulem}
\usepackage[varg]{txfonts}
\usepackage{multirow}
\usepackage[colorlinks, pdfborder={0 0 0}]{hyperref}
\definecolor{LinkColor}{rgb}{0.75, 0, 0}
\definecolor{CiteColor}{rgb}{0, 0.5, 0.5}
\definecolor{UrlColor}{rgb}{0, 0, 0.75}
\hypersetup{linkcolor=LinkColor}
\hypersetup{citecolor=CiteColor}
\hypersetup{urlcolor=UrlColor}
\allowdisplaybreaks
\begin{document}

\newcommand{\FigStart}{\begin{figure}[h]}
\newcommand{\FigEnd}{\end{figure}}
\newcommand{\LNh}{\hat{\mathbf{L}}_\text{N}}
\newcommand{\LN}{\mathbf{L}_\text{N}}
\newcommand{\bS}{\mathbf{S}}
\newcommand{\bJ}{\mathbf{J}}
\newcommand{\e}{\mathrm{e}}
\newcommand{\rmi}{\mathrm{i}}
\newcommand{\flow}{f_0}
\newcommand{\fcut}{f_\mathrm{cut}}
\newcommand{\bchi}{\bm{\chi}}
\newcommand{\blambda}{\bm{\lambda}}
\newcommand{\bLambda}{\bm{\Lambda}}
\newcommand{\bchia}{\bm{\chi}_\text{a}}
\newcommand{\bchis}{\bm{\chi}_\text{s}}
\newcommand{\chis}{\chi_\text{s}}
\newcommand{\chia}{\chi_\text{a}}
\newcommand{\chiadL}{\bchia \cdot \LNh}
\newcommand{\chisdL}{\bchis \cdot \LNh}
\newcommand{\chisSqr}{\bchis^2}
\newcommand{\chiaSqr}{\bchia^2}
\newcommand{\chisDchia}{\bchis \cdot \bchia}
\newcommand{\cA}{\mathcal{A}}
\newcommand{\cB}{\mathcal{B}}
\newcommand{\cC}{\mathcal{C}}
\newcommand{\cP}{\mathcal{P}}
\newcommand{\Mc}{M_\mathrm{c}}
\newcommand{\thetaz}{\theta_{0}}
\newcommand{\thetat}{\theta_{3}}
\newcommand{\thetats}{\theta_\mathrm{3S}}
\newcommand{\psiL}{\psi^\text{L}}
\newcommand{\dphi}{\partial \psi}
\newcommand{\dtheta}{\partial \theta}
\newcommand{\dphiL}{\partial \psi^\text{L}}
\newcommand{\derb}{\partial_b}
\newcommand{\dera}{\partial_a}
\newcommand{\df}{{\mathrm{d}f}}
\newcommand{\match}{\mathcal{M}}
\newcommand{\bOmega}{\mathbf{\Omega}}
\newcommand{\btheta}{\bm{\theta}}
\newcommand{\SBank}{\textsc{SBank}}
\newcommand{\LALSuite}{\textsc{LALSuite}}
\newcommand{\FFe}{\mathrm{FF_{eff}}}
\newcommand{\FF}{\mathrm{FF}}
\newcommand{\bg}{\mathbf{g}}
\newcommand{\Mchirp}{\mathcal{M}_\text{c}}
\newcommand{\MM}{\mathcal{M}}
\newcommand{\blue}{\color{blue}}
\newcommand{\magenta}{\color{magenta}}
\newcommand{\red}{\color{red}}
\newcommand{\green}{\color{green}}
\newcommand{\imrphenb}{\textsc{IMRPhenomB}}    
\newcommand{\eobnr}{\textsc{EOBNRv2}} 
\newcommand{\SqrtMet}{$\sqrt{|\textbf{g}|}$ } 
\newcommand{\msun}{$M_{\odot} $}

\title{Testing the Kerr nature of supermassive and intermediate-mass black hole binaries using spin-induced multipole moment measurements}

\date{\today} 
\author{N. V. Krishnendu} \email{krishnendu@cmi.ac.in}
\affiliation{Chennai Mathematical Institute, Siruseri, 603103, India}
\affiliation{Max Planck Institute for Gravitational Physics (Albert Einstein Institute), Callinstr. 38, 30167 Hannover, Germany}
\author{Anjali B. Yelikar} \email{anjaliy@cmi.ac.in}
\affiliation{Chennai Mathematical Institute, Siruseri, 603103, India}
\affiliation{Center for Computational Relativity and Gravitation, Rochester Institute of Technology, Rochester, New York 14623, USA}
\begin{abstract}
  The gravitational wave measurements of spin-induced multipole moment coefficients of a binary black hole system can be used to distinguish black holes from other compact objects~\cite{Krishnendu:2017shb}.  Here, we apply the idea proposed in Ref.~\cite{Krishnendu:2017shb} to binary systems composed of supermassive and intermediate-mass black holes and derive the expected bounds on their Kerr nature using future space-based gravitational wave detectors. Using astrophysical models of binary black hole population, we study the measurability of the spin-induced quadrupole and octupole moment coefficients using LISA and DECIGO. The errors on spin-induced quadrupole moment parameter of the binary system are found to be {  $\leq 0.1$ for almost $3\%$ of the total supermassive binary black hole population which is detectable by LISA whereas it is $\sim 46\%$ for the intermediate-mass black hole binaries observable by DECIGO at its design sensitivity.} We find that { errors on} {\it both} the quadrupole and octupole moment parameters can be estimated to { be} $\leq 1$ for $\sim 2\%$  and $\sim 50\%$ {of the population} respectively for LISA and DECIGO detectors. { Our findings suggest that  a subpopulation of binary black hole events, with the signal to noise ratio thresholds greater than 200 and 100 respectively for LISA and DECIGO detectors, would permit tests of black hole nature to 10\% precision.} 
  
 \end{abstract}

\pacs{} \preprint{} \maketitle

\section{Introduction}\label{intro}
        The detection of binary black hole mergers by Laser Interferometric Gravitational wave Observatory (LIGO)  and Virgo detectors~\cite{TheLIGOScientific:2014jea,TheVirgo:2014hva} have firmly established the existence of stellar mass black holes~\cite{Discovery,GW151226,GW170104,LIGOScientific:2018mvr,LIGOScientific:2018jsj,O1BBH,GW170608,GW170814}. The masses of detected compact binary systems, in the source frame, roughly range between $\sim7-80$\msun~\cite{LIGOScientific:2018jsj}. 
         
         Various electromagnetic observations tell us that there exists a supermassive black hole (SMBH) of mass $\sim 10^{5}-10^{10}M_{\odot}$ at the center of each galaxy~\cite{Ferrarese:2004qr}. Sagittarius $A^{*}$ is the closest supermassive black hole situated at the center of our Milky Way galaxy with a mass of about $4\times 10^{6}M_{\odot}$~\cite{Boehle2016,Gillessen2017}. Observational evidence for supermassive black holes also include the quasar observations from the Sloan Digital Sky Survey~\cite{Fan2005_Quasars_SloanDigitalSurvey}, a recent study which combined information from the Sloan Digital Sky Survey, the Two Micron All Sky Survey, and the Wide-field Infrared Survey Explorer~\cite{Xue-Bing2015Nature} and the quasar (ULAS J1120+0641) with mass $\sim2\times10^{9}M_{\odot}$~at a redshift of $\sim 7.085$ identified by the United Kingdom Infrared Telescope (UKIRT) Infrared Deep Sky Survey (UKIDSS) Eighth Data Release in 2010~\cite{Mortlock2011Nature},  to name a few. The formation mechanism of such systems is still not completely understood, though they are proposed to have formed through galaxy mergers~\cite{MassiveBHbinariesAGNsBegelman1980Nature}.  The first-ever image of a supermassive black hole situated at the centre of the elliptic galaxy  M87  has been produced by the Event Horizon Telescope (EHT) team. This radio source is situated around 16 Mpc away with a mass of $\sim6.5\times10^{9}M_{\odot}$~\cite{EHTpaper1_2019ApJ}.

        The mass gap between stellar-mass and supermassive black holes is expected to be filled by intermediate-mass black holes (IMBHs) having masses in the range of $\sim10^2-10^5$\msun~\cite{Mezcua_2017,vanderMarel:2003ka}. Indirect evidence for IMBHs from electromagnetic observations is promising and also motivate new proposals for gravitational wave detectors in the corresponding frequency range. The ultraluminous X-ray source HLX-1 hosted by galaxy ESO 243-49 is believed to be an intermediate-mass BH of mass $\sim 500 M_{{\odot}}$~\cite{Farrelletal2009Nature}. Another observational evidence for intermediate-mass BH came from the X-ray quasi-periodic oscillations of M82 X-1, which is the brightest X-ray source in the galaxy M82~\cite{Pasham2015tNature}. In Ref.~\cite{BulentKzltan2017Nature}, authors demonstrated the existence of an electromagnetically dark black hole in the globular cluster 47 Tucanae with mass $\sim2300 M_{\odot}$ through the observed pulsar acceleration rates together with N-body simulations, {though this claim is disputed in Ref.~\cite{Mann:2018xkm}}.  
  
        From a fundamental physics standpoint, one would like to understand how consistent these observations are with Kerr black holes of general relativity (GR).    The detected GW events till date are in agreement with general relativity as verified by several tests \cite{TOG,Meidam:2014jpa,Yunes:2013dva,AIQS06a,AIQS06b,Berti:2018cxi, Berti:2018vdi,Will:1977wq,TIGER2014,Yunes:2009ke,Will:1997bb,Samajdar:2017mka, Ghosh:2017gfp,BNSTGRAbbott2018,Isi:2019aib}. But we cannot  {\it fully} rule out the possibility of alternatives such as boson stars (BSs)~\cite{BosonStars}, gravastars (GSs)~\cite{gravastar_Mazur}, {\it etc} which can mimic the binary black hole signals~\cite{Giudice,Cardoso:2019rvt}.
        
         A novel method to test the binary black hole nature of the compact binary system to distinguish it from BH mimickers by measuring the spin-induced quadrupole moments was proposed in  Ref.~\cite{Krishnendu:2017shb}.   This method has been applied to the two inspiral-dominated events from first and second observational runs of LIGO and Virgo detectors, GW151226~\cite{GW151226} and GW170608~\cite{GW170608}, and constraints {were} obtained on the black hole nature of the detected gravitational-wave signals for the first time~\cite{Krishnendu:2019tjp}. Projected bounds on the Kerr nature of the binary system for various mass and spin configurations {were} demonstrated in the context of both ground and space-based GW detectors in Refs.~\cite{Krishnendu:2017shb,Krishnendu:2018nqa}.

        Spin-induced multipole moments arise due to the spinning motion of the compact object and was first introduced in the context of inspiralling compact binary systems in Ref.~\cite{Poisson:1997ha}. The leading-order effect is the spin-induced quadrupole moment which appears first at second post-Newtonian (2PN) order in the gravitational waveform along with quadratic spin terms. Spin-induced quadrupole moment coefficient for a compact object can be schematically represented as, $Q=-\mathcal{\kappa}\,\chi^{2}\,m^{3}$ where $\mathcal{\kappa}=1$ for Kerr BHs and   $\mathcal{\kappa}\sim 2-14$~\cite{Laarakkers:1997hb,Pappas:2012qg,Pappas:2012ns} for neutron stars (NSs) and $\mathcal{\kappa}\sim 10-150$ for boson stars (BSs)~\cite{Ryan97b}. The next-to-leading-order contribution (spin-induced octupole moment parameter) is a 3.5PN effect and appears with cubic spin terms in the phasing formula which can be schematically written as, $O=-\mathcal{\lambda}\,\chi^{3}\,m^{4}$ with $\mathcal{\lambda}=1$ for Kerr BHs, $\mathcal{\lambda}\sim 4-30$~\cite{Laarakkers:1997hb,Pappas:2012qg,Pappas:2012ns} for NSs, and $\mathcal{\lambda}\sim 10-200$ for BSs~\cite{Ryan97b}. We define the symmetric combinations of the spin-induced quadrupole and octupole moment coefficients as, $\mathcal{\kappa}_s=\frac{1}{2}(\mathcal{\kappa}_1+\mathcal{\kappa}_2)$ and $\mathcal{\lambda}_s=\frac{1}{2}(\mathcal{\lambda}_1+\mathcal{\lambda}_2)$, where the subscripts 1 and 2 denote the two objects in the compact binary system. The values of $\mathcal{\kappa}_s$ and $\mathcal{\lambda}_s$ are { 1}  if the binary system is composed of two black holes. Here we address the possibility of measuring spin-induced quadrupole and octupole moment coefficients of supermassive and intermediate-mass binary black holes using the proposed space-based gravitational wave observatories.
        
        Ground-based second-generation GW detectors started their third observation with improved sensitivity~\cite{AdvancedLIGO,TheVirgostatus} { in 2019}. There are proposals for third-generation ground-based GW detectors with enhanced sensitivity such as Einstein Telescope (ET) \cite{Sathyaprakash:2011bh} and Cosmic Explorer (CE) \cite{Regimbau:2012ir, Hild:2010id, Hild:2008ng,Evans:2016mbw}. Third-generation GW detectors can probe up to 1 Hz unlike the case of second-generation detectors where the sensitivity is not good {for frequencies lesser} than $\sim10$ Hz.
    
        The sensitivity of ground-based gravitational-wave detectors at lower frequencies is limited by the seismic noise~\cite{HughesThorne1998PRD,PeterSaulson1984PRD}. To overcome this and to extend the gravitational wave frequency spectrum to even lower frequencies we need detectors which operate at frequencies less than 1 Hz~\cite{BBOHarry2006,Amaro_Seoane_2012_eLISA_NGO,LISApathfinder2019March,DecidoPathFinder2009,Isoyama2018_DECIGO-B}. The space-based gravitational wave detectors,  such as Laser Interferometric Space Antenna (LISA)~\cite{LPF2017,LPFFirstResultsPRL,LISApathfinder2019March,amaroseoane2017laser}, DECi-hertz Interferometer Gravitational-wave Observatory (DECIGO)~\cite{DECIGOProposalSetoNakamura2001, DECIGOstatus2017,DecidoPathFinder2009,Isoyama2018_DECIGO-B} and Big Bang Observer (BBO)~\cite{BBOHarry2006,Cutler:2005qq}, will have the capability to probe gravitational wave frequencies from a few milli-Hz to tens of Hz. Among these, the LISA configuration is already funded and is expected to be operational by 2034 after successfully demonstrating some of the key technologies it will use, through the LISA Pathfinder mission~\cite{LPF2017,LPFFirstResultsPRL,LISApathfinder2019March} which was launched in December 2015. 
         
         The Japanese DECIGO mission is designed to bridge the gap between terrestrial GW detectors and LISA and is expected to operate in the deci-Hz band~\cite{DECIGOProposalSetoNakamura2001, DECIGOstatus2017,DecidoPathFinder2009,Isoyama2018_DECIGO-B}.  
         Though the DECIGO configuration was initially designed to probe signatures of the early universe including cosmic acceleration and gravitational wave background from inflation~\cite{DECIGOProposalSetoNakamura2001}, one can also look for intermediate-mass black hole binaries with masses of the order of a few hundred-to-thousand solar masses~\cite{Yagi:2012gb}, along with binary neutron stars and stellar-mass binary black holes~\cite{YagiSetoNaoki2011,Yagi:2013du}. Currently, the DECIGO configuration is a proposal whose science potential is being assessed~\cite{DECIGOstatus2017}.

\begin{figure*}[hbtp]
    \includegraphics[scale=0.45]{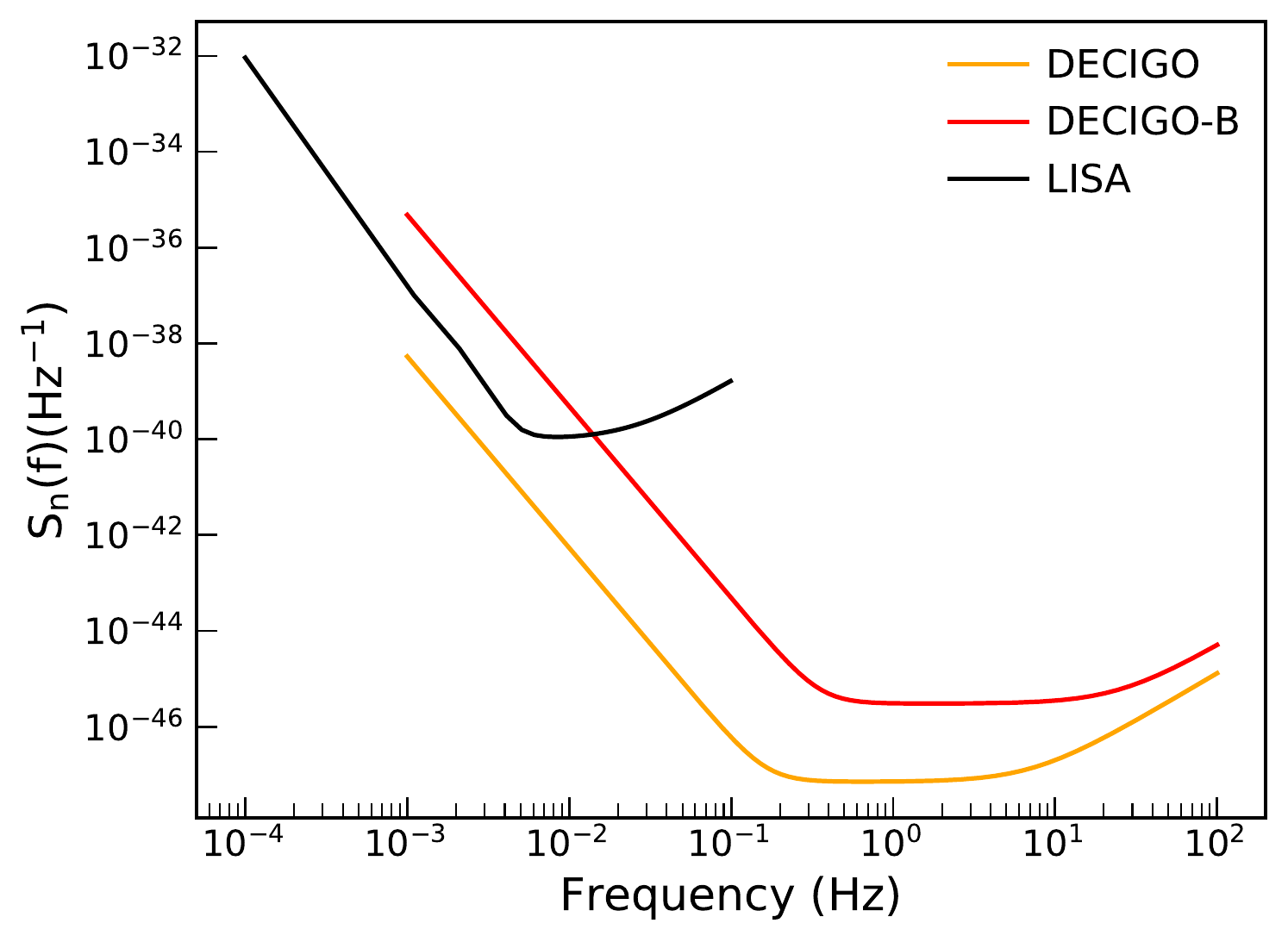}
    \includegraphics[scale=0.45]{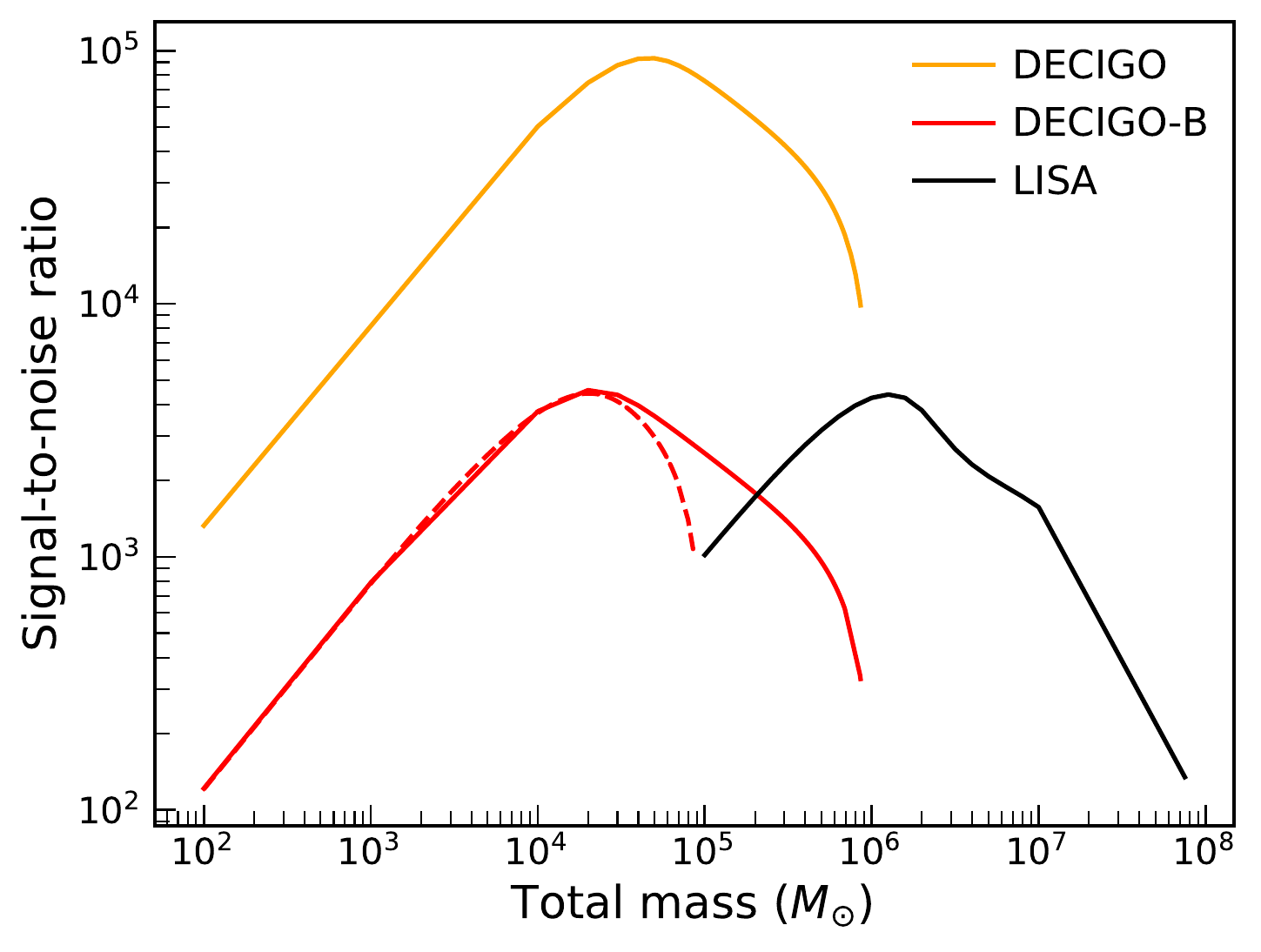}
    \caption{Left panel: Noise PSDs of LISA~\cite{Babak2017} (black), DECIGO~\cite{YagiSetoNaoki2011} (orange)  and DECIGO-B~\cite{Isoyama2018_DECIGO-B} (red) configurations. Right panel: Corresponding signal-to-noise ratios (SNR) for LISA (black), DECIGO (orange)  and DECIGO-B (red) configurations as a function of total mass of the binary system. In order to obtain the SNR values we assume the binary to be optimally oriented at a luminosity distance of 1 Gpc. In the case of basic DECIGO configuration (DECIGO-B), the SNR is plotted considering two different lower cut-off frequencies, $\rm{f_{min}}=10^{-1}$Hz ($f_{1}$, dashed curve) and $\rm{f_{min}}=10^{-2}$Hz ($f_{2}$, solid curve).}
    \label{fig:NoiseCurves}
    \end{figure*}

        In this paper, we investigate the measurability of spin-induced quadrupole and octupole moment parameters of supermassive and intermediate-mass binary black holes using LISA and DECIGO detector configurations, respectively. Further, we show that the proposed LISA and DECIGO detectors will allow us to measure both spin-induced quadrupole and octupole moment parameters with reasonably good statistical errors and hence are excellent probes for the tests of Kerr nature of compact binary systems composed of supermassive and intermediate-mass binary black holes, by considering an astrophysical population of binary black holes. { We show our main results in Table~\ref{Tab:MainResults}. The numbers in Table~\ref{Tab:MainResults} correspond to the percentage of sources that satisfy the detection threshold of LISA and DECIGO and errors on spin-induced multipole moment parameters less than a certain value (see Sec.~\ref{subsec:population} for more details). From Table~\ref{Tab:MainResults}, it is evident that the gravitational wave observations of spin-induced multipole moment parameters  can give stringent constraints on the allowed parameter space of black hole mimickers such as the spinning boson star models in Ref.~\cite{Ryan97b}.}

    \begin{table}
    \begin{center}
        \begin{tabular}{lcccc}
            \hline\hline
            \addlinespace[2mm]
            \multirow{1}{*}{Detector} & $\Delta\kappa_{s}\leq1$ \hspace{12mm}&$\Delta\kappa_{s}\leq0.1$ \hspace{12mm}&$\Delta\kappa_{s}\leq1$ and $\Delta\lambda_{s}\leq1$  \\
            \addlinespace[1mm]
            \hline\hline
            \addlinespace[1mm]
            LISA& \multirow{2}{*}{53.48} &  \multirow{2}{*}{2.30 } &   \multirow{2}{*}{1.09}\\
            &  &  &  &  \\
            \addlinespace[1mm]
            \hline
            \addlinespace[1mm]
            DECIGO& \multirow{2}{*}{90.13} &  \multirow{2}{*}{45.99}  &   \multirow{2}{*}{49.97}\\
            &  &  & & \\
            \addlinespace[1mm]
            \hline\hline
        \end{tabular}
        \caption{{ The numbers correspond to a fraction of total population (in percentage) of the binary systems which give errors on spin-induced quadrupole and octupole moment parameters better than a certain accuracy from the binary black hole simulations of supermassive and intermediate-mass binary black holes  described in Sec.~\ref{subsec:population}}.}
        \label{Tab:MainResults}
    \end{center}
    \end{table}


        We start with a brief description of the method and the parameter estimation technique we use for this analysis (Sec.~\ref{sec:details}). In Sec.~\ref{sec:res}, we detail the main results obtained from our study and summarise our findings in Sec.~\ref{sec:con}.

\section{Details of the analysis}
\label{sec:details}


\begin{figure}[hbtp]
    \includegraphics[scale=0.45]{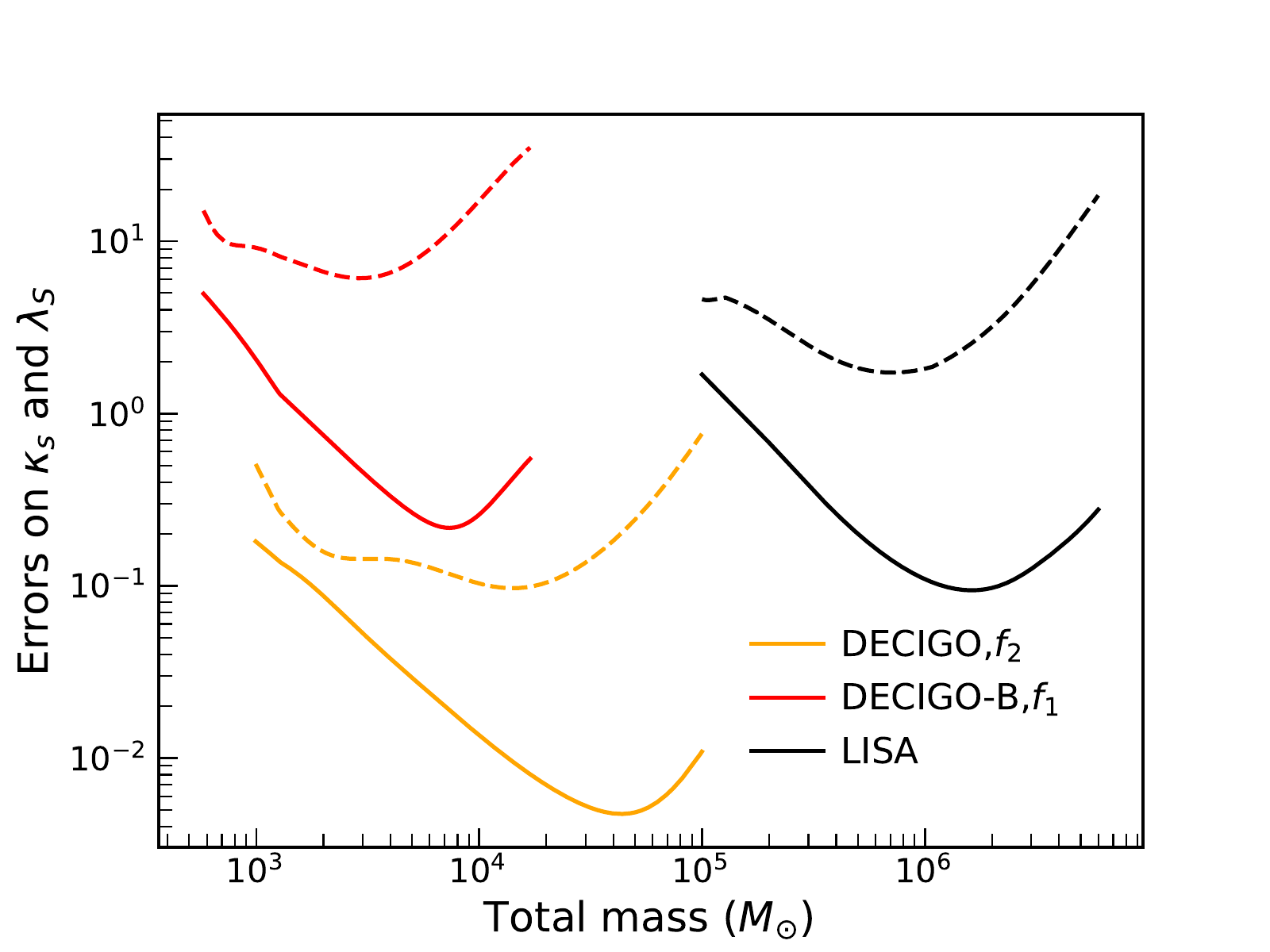}
    \caption{Errors on spin-induced quadrupole moment parameter ($\Delta\mathcal{\kappa}_{s}$, solid curves) and octupole moment parameter ($\Delta\mathcal{\lambda}_{s}$, dashed curves) as a function of the total mass of the binary system which is assumed to be located at a luminosity distance of 1 Gpc {and oriented with a particular configuration in the sky} with spin magnitudes (0.6, 0.3) and the mass ratio of 1.1. Black, orange and red curves respectively show the results obtained when we consider LISA, DECIGO and DECIGO-B configurations. We assume $\rm{4\, yr}$ of observation time for all the three detector configurations.}
    \label{fig:Mq}
    \end{figure}

\subsection{Waveform model}
\label{subsec:wf}

        The waveform model we employ is the same as that of Ref.~\cite{Krishnendu:2018nqa} that describes the inspiral dynamics of a non-precessing compact binary system which is based on the post-Newtonian (PN) technique \cite{Blanchet:2013haa,Marsat:2012fn, Bohe:2012mr,
Bohe:2013cla, Marsat:2013caa, Bohe:2015ana, Marsat:2014xea, Arun:2008kb,
Kidder:1995zr,Will:1996zj, Buonanno:2012rv, Mishra:2016whh},   and can be schematically represented as, 
\begin{equation} \tilde{h}(f)=\frac{M^2}{D_L} \sqrt{\frac{5\,\pi
\,\eta}{48}}\sum_{n=0}^{4} V_2^{n-7/2}\,C_{2}^{(n)}\,\mathrm{e}^{\mathrm{i}
\,\big({2\,\Psi}_\mathrm{SPA}(f/2)-\pi/4\big)}\, , 
\label{eq:waveform1}
\end{equation}
where $\Psi_\mathrm{SPA}(f)$ is the 4PN \footnote{At 4PN only the spin-orbit terms are available.} accurate point particle phase which explicitly contain spin-induced quadrupole moment terms at 2PN, 3PN, and 3.5PN and leading order spin-induced octupole moment term at 3.5PN. The leading (second) harmonic and its corrections are incorporated up to 2PN (n=4) order in the amplitude and appear through the coefficients
$C_{2}^{(n)}$~\cite{Arun:2008kb}. 
In Eq.~\ref{eq:waveform1},  $M$, $\eta$ and $D_L$ denote the total mass, the symmetric mass ratio of the binary system and the luminosity distance to the source. The pre-factor  $V_{2}$ is a function of the total mass of the binary system and the gravitational wave frequency (See Sec. 2 of~\cite{Krishnendu:2018nqa} and supplemental material of~\cite{Krishnendu:2017shb} for more details about the waveform). Notice that the inclusion of precession effects in the waveform may lead to tighter constraints because of the additional features in the precessing waveforms. However, due to the unavailability of analytical precessing waveform models, we postpone this for  future work.

\subsection{Parameter estimation and detector configurations}
         { We use Fisher information matrix analysis to obtain the measurement errors on spin-induced quadrupole moment parameters of supermassive and intermediate-mass black holes. Fisher information matrix approach ~\cite{Vallisneri:2007ev,Krishnendu:2018nqa,Weinstein_Weiss_1988}  is a semi-analytical parameter estimation technique
which can be used to compute the $1-\sigma$ error bars on parameters characterizing the gravitational wave signal, given a waveform model  and the detector sensitivity.}  The elements of the matrix are defined as follows,
\begin{equation}
 \Gamma_{ij}=2\int_{f_{\rm{lower}}}^{f_{\rm{upper}}}\,\left(\partial_{i}\tilde{h}(f)\,\partial_{j}\tilde{h}^{\ast}(f)+\partial_{i}\tilde{h}^{\ast}(f)\,\partial_{j}\tilde{h}(f)\right)\frac{df}{S_{n}(f)}\,,
\label{eqn:Fisher}
\end{equation}
where we denote the frequency domain gravitational waveform as $\tilde{h}(f)$ and its partial derivative with respect to the $i^{th}$ parameter of the binary system as $\partial_{i}\tilde{h}(f)$ (here, $\partial_{i}\tilde{h}^{\ast}(f)$ is the conjugate of $\partial_{i}\tilde{h}(f)$).  In our case the set of parameters in the signal manifold which characterise the compact binary consists of,
\begin{equation}
    \overrightarrow{\theta}=\bigg\{ {\mathcal{M}}, \,\delta, \,\chi_{1},\,\chi_{2},\,\mathcal{\kappa}_{s},\,                \mathcal{\lambda}_{s}, \, t_{c},\,\phi_{c}\bigg\},
    \label{eqn:parameterspace}
    \end{equation}
    where $t_{c}$, $\phi_{c}$ are the time and phase at coalescence and ${\mathcal{M}}$ and $\delta$ are the chirp mass and the asymmetric mass ratio of the system 
     and $\chi_{1}$, $\chi_{2}$ are the magnitudes of dimensionless spin parameters. 
{ The symmetric combination of the} spin-induced quadrupole ($\kappa_1$ and $\kappa_2$) and octupole moment parameters ($\lambda_1$ and $\lambda_2$) of the binary system are denoted by $\mathcal{\kappa}_s$ and $\mathcal{\lambda}_s$, respectively.  The $1-\sigma$ error bars on each parameter (Eq.~(\ref{eqn:parameterspace})) are computed in the high signal-to-noise ratio (SNR) limit~\cite{Vallisneri:2007ev,Yunes:2016jcc} as $\Delta\overrightarrow{\theta}=\sqrt{\Gamma_{ii}^{-1}}$, under the assumption that the detector noise is stationary and Gaussian. Notice that while estimating the errors on  $\kappa_{s}$ and  $\lambda_s$, we set the anti-symmetric combination to be zero, {\it{i.e., $\mathcal{\kappa}_a=\mathcal{\lambda}_a=0$}}.

The bounds we obtained from the Fisher matrix analysis may be seen as a typical order of magnitude estimates and a detailed study based on Bayesian analysis is required to make a more precise quantification of the same. As the Fisher matrix estimates are expected to agree with those from a numerical sampling of the likelihood in the large SNR limit, we have considered systems that have SNRs of the order of hundreds~\cite{OShaughnessy:2013zfw,Cokelaer_2008,Vallisneri:2007ev,Vitale:2010mr}.  Further, we make sure that the Fisher matrices used in the analysis are not ill-conditioned and discard those which show numerical issues during inversion.

Gravitational wave detector noise is characterised by the noise power spectral density, $S_{n}(f)$ in Eq.~(\ref{eqn:Fisher}).        
  The noise spectral density used for LISA is given by~\cite{Babak2017},
 \begin{widetext}
\begin{equation}
S_{n}(f)=\frac{20}{3\,L^2}\Big(4\,S_{n}^{acc}(f)+2\,S_{n}^{loc}+S_{n}^{sn}+S_{n}^{omn}\Big)\\ \,\Bigg[1+\bigg(\frac{2\,L\,f}{0.41c}\bigg)^{2}\Bigg] +S_{n}^{gal},
\label{eqn:PSDLISA}
\end{equation}
where,
\begin{eqnarray}
S_{n}^{acc}&=& \Bigg\{9\times10^{-30} + 
    3.24\times 10^{-28} \Bigg[\bigg(\frac{3\times10^{-5}\,\rm{Hz}}{f}\bigg)^{10}   \nonumber  \\
&&
+ \bigg(\frac{10^{-4}\,\rm{Hz}}{f}\bigg)^2\Bigg]\Bigg \}\,\frac{1}{(2\, \pi\,  f)^{4}}\,\rm{m^2\,Hz^{-1}}, \nonumber \\
S_{n}^{loc}&=& 2.89\times10^{-24}\,{\rm{m^2\,Hz^{-1}}}, \nonumber \\
S_{n}^{sn}&=&7.92\times10^{-23}\,{\rm{m^2\,Hz^{-1}}},\nonumber \\
S_{n}^{omn}&=&4.00\times10^{-24}\,{\rm{m^2\,Hz^{-1}}}, \nonumber\\
S_{n}^{gal}&=&1.633\times10^{-44}\,\bigg(\frac{f}{1\,\rm{Hz}}\bigg)^{-7/3}\mathrm{\exp}\left({-\bigg(\frac{f} {1.426\,\rm{mHz}}\bigg)^{1.183}}\right)\,\left(1+{\tanh \bigg(-\frac{f-2.412\,\rm{mHz}}{4.835\,\rm{mHz}}\bigg)}\right)\,\rm{Hz^{-1}}. \nonumber 
\end{eqnarray}
\end{widetext}

Here $S_{n}^{acc}$, $S_{n}^{loc}$, $S_{n}^{sn}$, $S_{n}^{omn}$ and $S_{n}^{gal}$ are due to low-frequency acceleration, local interferometer noise, shot noise, other measurement noise, and galactic confusion noise, respectively. The detector arm length $L$ is fixed to be $2.5\times10^{9}$meters and $c$ is the speed of light in meters per second.
Noise spectral densities used for DECIGO~\cite{YagiSetoNaoki2011} and DECIGO-B~\cite{Isoyama2018_DECIGO-B} are,

\begin{widetext}  
\begin{eqnarray}
S_{n}(f)=7.05\times10^{-48} \left(1 + \bigg(\frac{f}{ 7.36\,{\rm{Hz}}}\bigg)^2\right) + 4.8\times10^{-51}\, \bigg(\frac{f}{1\,\rm{Hz}}\bigg)^{-4 }\Bigg(1 +\bigg (\frac{f}{ 7.36\,{\rm{Hz}}}\bigg)^{2}\Bigg)^{-1} + 5.33\times10^{-52 }\,\bigg(\frac{f}{1\,\rm{Hz}}\bigg)^{-4}\,\rm{Hz^{-1}},
\label{eqn:DECIGO-PSD}
\end{eqnarray}
and
\begin{eqnarray}
S_{n}(f) =3.03 \times 10^{-46}\,\left(1 + 1.584\times10^{-2}\,\left(\frac{f}{1\,{\rm{Hz}}}\right)^{-4} + 1.584\times 10^{-3}\left(\frac{f}{1\,{\rm{Hz}}}\right)^{2}\right)\,\rm{Hz^{-1}},
\label{eqn:DECIGOB-PSD}
\end{eqnarray}
\end{widetext}
respectively.

        The lower and upper cut-off frequencies (see Eq.~(\ref{eqn:Fisher})) for the analysis are fixed using the following relations,
         \begin{align}
         f_{\rm {upper}}&=\rm{min}\big(f_{\rm{max}}, \,f_{\rm{ISCO}}\big) \nonumber \\
         f_{\rm {lower}}&=\rm{max}\big(f_{\rm{min}},\,f_{\rm{4 \,yr}} \big).
         \label{eqncutoffs}
         \end{align}
For LISA, we fix $f_{\rm{max}}$ as $0.1$ Hz and $f_{\rm{min}}$ to be  $10^{-4}$ Hz and for  DECIGO ${f_{\rm{max}}}$ is taken to be 10 Hz throughout the study. To examine the effect of the lower cut-off frequency on different DECIGO configurations, we consider two different scenarios, basic DECIGO or DECIGO-B~\cite{Isoyama2018_DECIGO-B,DECIGOProposalSetoNakamura2001} with a conservative low-frequency cut-off of $10^{-1}$Hz{$(f_{1})$} and DECIGO~\cite{YagiSetoNaoki2011} at its designed sensitivity (which we refer to as DECIGO) with a low-frequency cut-off of  $10^{-2}$Hz{$(f_{2})$}. As expected, due to the improved sensitivity, bounds obtained from DECIGO are much better than DECIGO-B in general.  For the current analysis, the waveform model we use has information about the spin-induced multipole moment parameters only in the inspiral as described in Sec.~\ref{subsec:wf}. By truncating the analysis at $f_{\rm{upper}}$, given in Eq.~\ref{eqncutoffs}, we avoid any systematic biases that might arise due to the presence of merger-ringdown phases of the dynamics in the waveform. { To obtain the value of $f_{\rm{ISCO}}$, which corresponds to the innermost stable circular orbit (ISCO) frequency of Kerr BHs, we use a fitting formula which is a function of the masses and spins of the binary constituents~\cite{Husa:2015iqa,EccPEFavata,Krishnendu:2017shb,Krishnendu:2018nqa}.  }
Our choice of lower cut-off frequency accounts for the fact that the compact binary system spends four years in each of the detector bands. In order to achieve this, we take $f_{\rm 4\,yr}=f_{\rm{upper}}\left(1 + \frac{m_{1}\, m_{2}}{(m_{1}+m_{2})^3}\, \tiny{6.6\times10^4}\,\rm{T^{-1}} \right)^{-\frac{3}{8}} $
, where $\rm{T}$ is fixed to be $\rm{4 \,yrs}$~\cite{AISS2007}. 

{ Left panel of  Fig.~\ref{fig:NoiseCurves} shows the noise PSDs of LISA (black), DECIGO (orange) and DECIGO-B (red) configurations, whereas the right panel shows the variation of signal-to-noise ratios (SNR) with the total mass of the compact binaries in the respective frequency bands. The signal-to-noise ratios are calculated by assuming binary systems optimally oriented at a luminosity distance of 1 Gpc with a mass ratio of 1.1 and dimensionless component spins (0.6, 0.3). Notice the improvement in the signal-to-noise ratios at the high mass end when we choose lower cut-off frequency a factor less ($\rm{f_{low}}=10^{-2}$Hz, red solid curve) than the conservative value  ($\rm{f_{low}}=10^{-1}$Hz, red dashed curve) in the case of DECIGO configurations.}

\begin{figure*}[hbtp]
\includegraphics[scale=0.45]{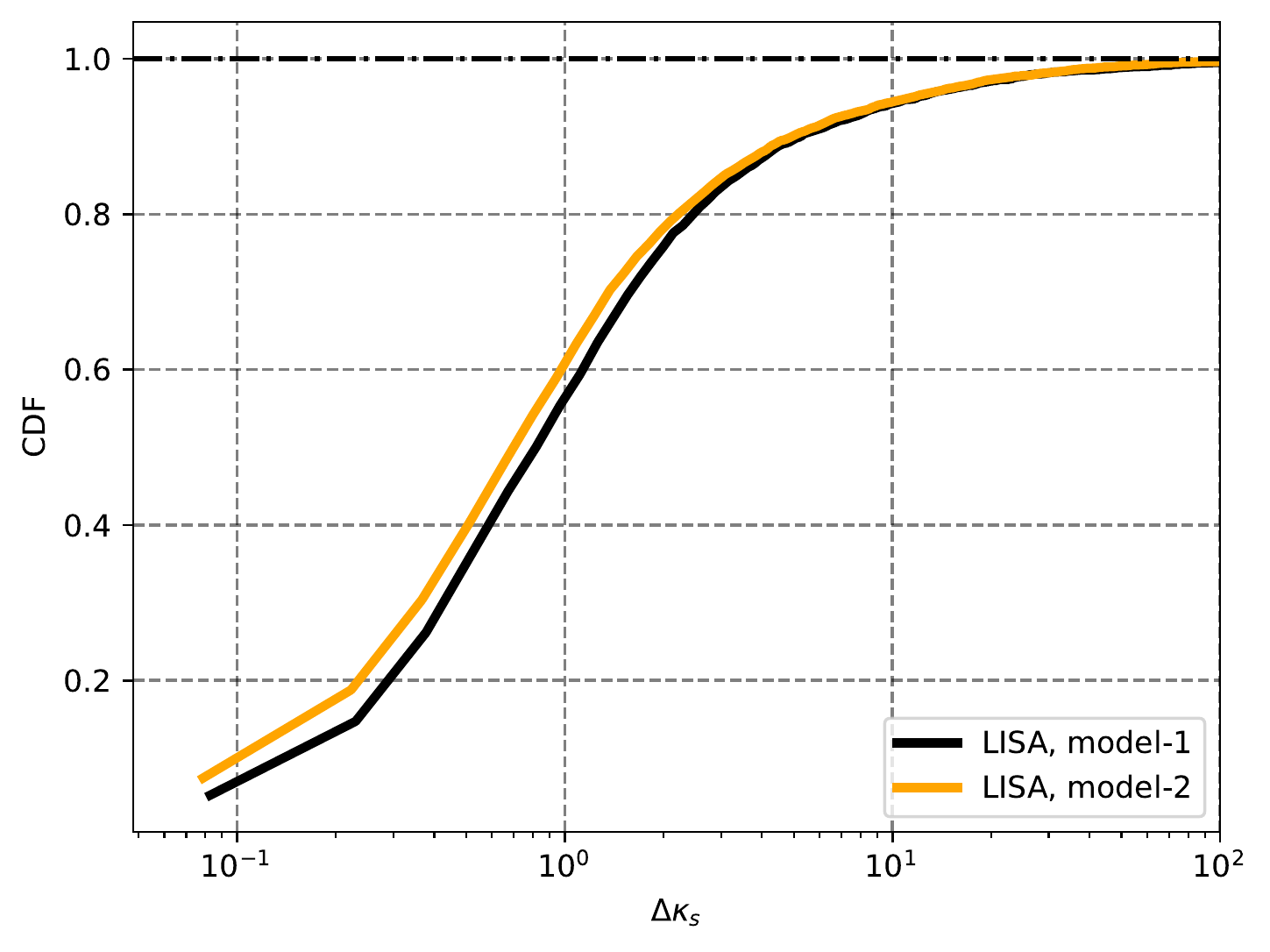}
\includegraphics[scale=0.45]{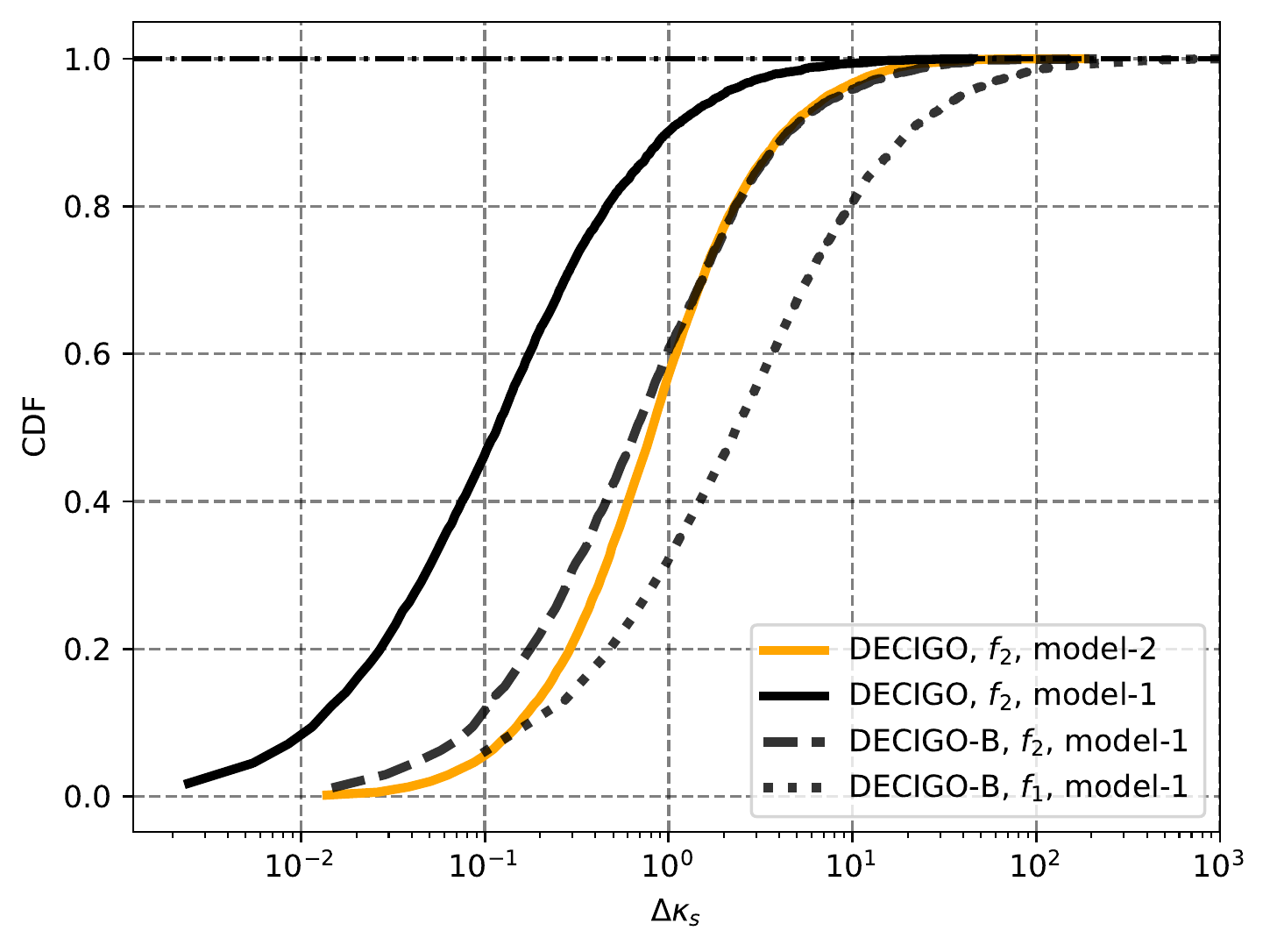}
\caption{Cumulative distributions of errors on the spin-induced quadrupole moment coefficients for an astrophysical population of compact binary systems using LISA and DECIGO detectors. We choose to show results from two models, model-1(black curve) and model-2 (orange curve), for both detectors as described in the text (see Sec.~\ref{sec:res} for more details).  Two detector configurations of DECIGO,  DECIGO-B, and DECIGO are considered. For DECIGO-B configuration, results obtained from two different lower cut-off frequencies $10^{-1}$Hz ($f_{1}$, black dotted curve) and $10^{-2}$Hz ($f_{2}$, black dashed curve) are compared assuming model-1.}  
\label{Fig:Population_Ks}
\end{figure*}

\begin{figure}[hbtp]
\includegraphics[scale=0.45]{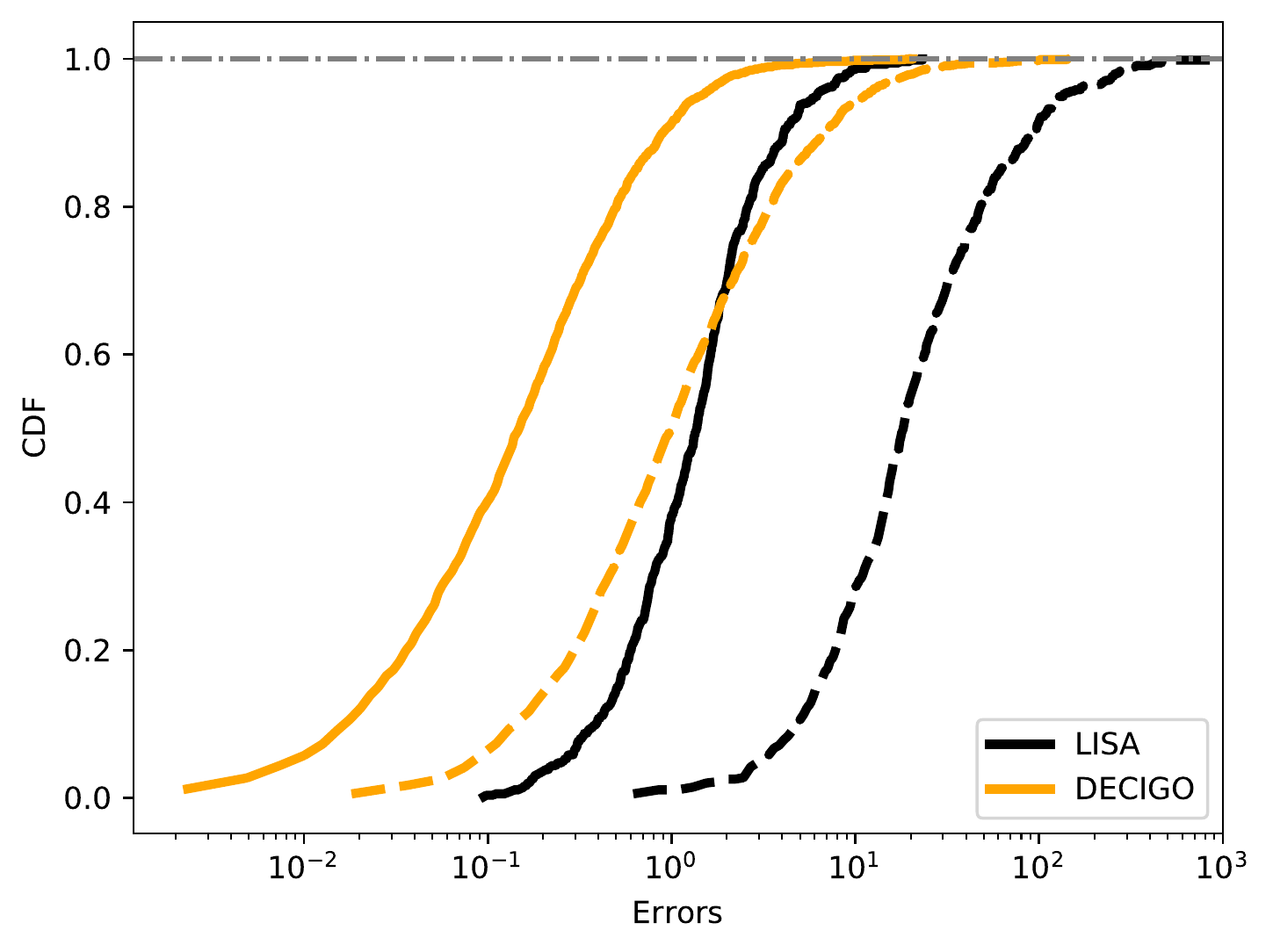}
\caption{Cumulative distributions of errors on the spin-induced quadrupole (solid lines) and octupole moment  (dashed lines) coefficients for an astrophysical population of compact binary systems, assuming model-1 for LISA (black, Q3-nod) and DECIGO (orange, model-1) configurations. } 
\label{Fig:Population_KsLs}
\end{figure}

\section{Testing the nature of supermassive and intermediate-mass binary black holes}
\label{sec:res}

        We consider two scenarios to demonstrate the method of testing the binary black hole nature of intermediate-mass and supermassive binary black holes. Firstly, we obtain the errors on spin-induced quadrupole and octupole moment parameters of the compact binary as a function of the total mass of the system keeping the spin magnitudes, mass ratio, location, and orientation fixed (Sec.~\ref{subsec:totalmass}).   Secondly, we investigate the applicability of this test for an astrophysical population of supermassive and intermediate-mass binary black holes assuming certain distributions for the source parameters (Sec.~\ref{subsec:population}). 

\subsection{Errors as a function of the total mass of the binary system}
\label{subsec:totalmass}

     In Fig.~\ref{fig:Mq}, we show the errors on spin-induced quadrupole (solid curve) and octupole moment (dashed curve) parameters as a function of the total mass of the system, where we fix the mass ratio $\left(m_{1}/m_{2}\right)$ to be 1.1 and the dimensionless component spins ($\chi_{1},\,\chi_{2}$) to be (0.6, 0.3).  Black curves in Fig.~\ref{fig:Mq} correspond to supermassive black holes (which LISA is more sensitive to) and, orange and red curves show the errors corresponding to intermediate-mass black holes (which DECIGO detectors will be more sensitive to).  As we can see from the figure, errors on both the parameters decrease as a function of total mass initially (irrespective of the detector configuration assumed) and then increase. This is because of the combined effect of SNR and the inspiral truncation frequency of the analysis. { As the total mass increases, the signal strength of a binary system with fixed location and orientation in the sky increases, but the upper cut-off frequency decreases as it is inversely related to the total mass, decreasing the number of cycles in the detector band.} 

\subsection{Errors from an astrophysical population of binary systems}
\label{subsec:population}

      The results shown in Fig.~\ref{fig:Mq} are not enough to completely assess the capabilities of LISA and DECIGO to carry out the test of Kerr nature of the binary black hole system, as { they} correspond to certain representative binary configurations. We repeat the analysis for a simulated population of binary black holes, that may be detected by LISA and DECIGO, and obtain the fraction of the total population this test would yield good constraints for.
      
       The simulated population in our case assumes that the binary black hole merger rate per redshift bin in the observer frame follows the relation,

\begin{eqnarray} 
\rm{\frac{d\,R(z)}{d\,z}}=\rm{\mathcal{R}}(z)\,\rm{\frac{d\,V_{c}(z)}{d\,z}},
\end{eqnarray}

    where $\rm{R}(z)$ is the number of binary coalescence per observation time, $\rm{\mathcal{R}}(z)$ is the merger rate density in the detector frame and $\rm{V_{{c}}}(z)$ denotes the comoving volume.

        For LISA sources, we assume the massive black hole rate evolution follows the models given in Klein et al.~\cite{Klein:2015hvg}. These semi-analytical massive black hole-galaxy coevolution models assume two different birth mechanisms for massive black holes and also account for the time delay between massive black hole merger and galaxy merger~\cite{Barausse:2012fy,Sesana:2014bea,Antonini:2015cqa,Antonini:2015sza,Bonetti:2018tpf,Volonteri:2002vz,Sesana:2008ur,Volonteri:2003tz}. Following the $\rm{R}(z)$ and detector frame total mass distributions given in Fig.~[3] of Ref.~\cite{Klein:2015hvg}, we populate binary black holes keeping the component masses nearly equal for three different formation mechanisms described in~\cite{Klein:2015hvg}\footnote{For Q3-nod (Q3-d) we populate up to a redshift of $z=19 (10)$ with total masses range between $8\times10^3-10^{8}M_{{\odot}} (2.21\times10^{4}-10^{8}M_{{\odot}})$.}.  
Among the three models, Model Q3-d (model-1), Model Q3-nod (model-2) and Model popIII (model-3), we observe that very few sources cross the detection threshold (SNR$\geq$200) for Model popIII and hence we only show the results obtained from Model Q3-d and Model Q3-nod here.

    In order to populate intermediate-mass black holes, which are interesting sources for the DECIGO configurations, we start with the following relation~\cite{LIGOScientific:2018jsj},
    \begin{eqnarray}
    \rm{\mathcal{R}}(z)=\rm{\mathcal{R}}_{0}\, \left(1+z \right)^{\lambda}\,\frac{\rm{T_{{obs}}}}{1+z},
    \end{eqnarray}

     where $\rm{\mathcal{R}}(z)$ is the rate of binary mergers that occur over the total observation time $\rm{T_{{obs}}}$ measured in the detector frame. Here, $\rm{\mathcal{R}}_{0}$ is a constant which gives the rate density corresponding to a particular value of redshift and we fix it to be 40 $\rm{Gpc^{-3} yr^{-1}}$. { The magnitude of $\rm{\mathcal{R}}_{0}$  will not affect our results as this is merely a scaling factor. } We distribute sources up to a redshift of 20 assuming two different population models for DECIGO configurations, model-1 and model-2. For  model-1, we fix $\lambda=0$ (rate density is assumed to be a constant with respect to the redshift) and the component masses to be uniformly distributed in the range $10^{2}-10^{4}M_{{\odot}}$.  
For  model-2, we fix $\lambda=6.5$~\cite{LIGOScientific:2018jsj} and the primary mass $(m_{1})$ is drawn from a power-law distribution with index $1.6$, in the range $10^{2}-10^{4}M_{{\odot}}$  and secondary mass $(m_{2})$ is drawn from a uniform distribution. The dimensionless spin parameters $\chi_{1}$ and $ \chi_{2}$ are distributed uniformly between -1 to 1 for both LISA and DECIGO sources.   Notice that, among the total populated sources positioned isotropically in orientation and polarisation sky, we choose only those sources which satisfy detection criteria set by the signal-to-noise ratios  200 and 100 for LISA and DECIGO/DECIGO-B respectively. We further perform PE on the signals which pass the signal-noise-ratio threshold using Fisher matrix analysis and the results are shown in Figs.~\ref{Fig:Population_Ks} and ~\ref{Fig:Population_KsLs}.

\subsubsection{ Constraints on the BBH nature from spin-induced quadrupole moment measurements}    The cumulative distribution of errors on spin-induced quadrupole moment parameters for an astrophysical population of supermassive (LISA sources) and intermediate-mass (DECIGO sources) black holes are shown in Fig.~\ref{Fig:Population_Ks}. For the case of LISA we show results from two models, model-1 (Q3-nod, black) and model-2 (Q3-d, orange). If we assume that the supermassive black hole binary formation mechanism is described by model-1, { $53.48\%$} of the total population provides errors on $\Delta\kappa_{s}$ $\leq$1. This changes to { $57.90\%$} if we consider model-2. As it is also clear from Fig.~\ref{Fig:Population_Ks} left panel, constraints on the spin-induced quadrupole moment parameters are not affected by the particular choice of astrophysical model for supermassive binary black holes. 

{ Right panel of Fig.~\ref{Fig:Population_Ks} shows estimates from different DECIGO configurations. We compare results from two astrophysical population models, model-1 (black) and model-2 (orange), for both DECIGO and DECIGO-B. To understand the effect of lower cut-off frequency, we show the bounds from two different lower cut-off frequencies $10^{-1}$Hz ($f_{1}$, black dotted curve) and $10^{-2}$Hz ($f_{2}$, black dashed curve) assuming model-1 and DECIGO-B configuration.  We find that $\Delta\kappa_{s}$ is measured with $10\%$ accuracy for { $45.99\%$} of intermediate-mass binary black hole population assuming model-1 using DECIGO configuration {at $10^{-2}$Hz lower cut-off frequency}. On the other hand, assuming model-2 we find that $\Delta\kappa_{s}$ $\leq$0.1 for { $5.08\%$} of the total population in the case of intermediate-mass black holes. Considering model-1 with a lower cut-off frequency of $10^{-1}$Hz,  { $2.68\%$}  of the total intermediate-mass binary population gives $\Delta\kappa_{s}$ $\leq 1$ using DECIGO-B and it increases to { $10.47\%$} when we use $10^{-2}$Hz as the lower cut-off frequency. }

\subsubsection{ Constraints on the BBH nature from spin-induced quadrupole and octupole moment parameters}    
Here we focus on the simultaneous measurability of spin-induced quadrupole and octupole moment coefficients for an astrophysical population of binary black holes. In Fig.~\ref{Fig:Population_KsLs}, we show errors on spin-induced quadrupole (solid lines) and octupole (dashed lines) measured using LISA (black) and DECIGO (orange) detectors. We restrict our analysis to model-1 for both supermassive and intermediate-mass binary black hole models.  

Among the simulated binaries which cross the LISA detection threshold, we find that { $1.09\%$} of the population has both $\Delta\kappa_{s}$ and $\Delta\mathcal{\lambda}_{s}$  $\leq$1. From the total population of intermediate-mass binary black holes detectable by DECIGO configuration with a lower cut-off frequency of $10^{-2}$Hz, $49.97\%$ of the sources give errors on both the spin-induced multipole moments $\leq$1 when we assume model-1. We conclude by noting that the spin-induced multipole moment coefficients of supermassive and intermediate-mass binary black holes, can be constrained well for a subpopulation of binary systems using LISA, DECIGO and DECIGO-B detectors, respectively.

\section{Summary and future directions}
\label{sec:con}

 	In this analysis, we investigated the possibility of testing the Kerr nature of intermediate-mass and supermassive binary black hole systems using space-based gravitational wave detectors. From the measurements of spin-induced multipole moment coefficients, we find that the space-based gravitational detectors DECIGO and LISA are excellent probes for the tests of Kerr nature of the compact binary systems composed of intermediate-mass and supermassive binary black holes respectively. Compared to the basic DECIGO (DECIGO-B) configuration with a conservative lower cut-off frequency, the performance of  DECIGO configuration is found to be improved for the entire parameter space.  
    The current analysis can be extended to test the black hole nature of the central object in extreme mass-ratio inspirals (EMRIs). The waveform development for EMRIs is still an open problem and an active field of research. We did not want to use the test particle limit of the PN waveforms to model EMRIs, which is known to miss several important physical effects. In short, we do not currently have the waveform models to study the test of BH nature of the central compact object in the case of EMRIs and hence we postpone this for future work.

{\it Acknowledgement:} N. V. K.  and A. B. Y thank K. G. Arun for useful discussions and support at each stage of the analysis. N. V. K.  and A. B. Y also thank M. Saleem and Shilpa Kastha for helpful discussions. N. V. K.  and A. B. Y are thankful to C. K. Mishra for critically reading the manuscript and giving comments. N. V. K is partially supported by a grant from Infosys Foundation. A. B. Y. acknowledges support by the grant EMR/2016/005594 by SERB.  We thank Chennai Mathematical Institute, Chennai and International Centre for Theoretical Sciences, Bangalore for giving access to their workstations.  

\bibliographystyle{apsrev}
\bibliography{lisadecigo}

\end{document}